\documentclass[epj]{svjour}
\usepackage{latexsym}
\usepackage{graphics,color}
\usepackage{bm} 
\begin{document}
\title{Interaction for the trapped fermi 
gas from a unitary transformation of the exact two-body
spectrum}
\author{J. Rotureau}                     
\institute{Fundamental Physics, Chalmers University of Technology, 
              412 96 G\"oteborg, Sweden}
\date{Received: date / Revised version: date}
%
\abstract{We study systems of few two-component fermions interacting in a Harmonic Oscillator trap. The fermion-fermion 
interaction is generated in a finite basis with a unitary transformation of the exact two-body 
spectrum given by the Busch formula. The few-body Schr\"odinger equation is solved with 
the formalism of the No-Core Shell Model. We present results for a system of three fermions interacting at unitarity 
as well as for finite values of the S-wave scattering length $a_2$ and effective range $r_2$. Unitary systems with
 four and five fermions are also considered.
We show that the many-body energies obtained in this approach are in excellent agreement with exact solutions for the three-body problem, and
results obtained by other methods in the other cases.
\PACS{
      {03.75.Ss} {Degenerate Fermi gases}   \and
      {34.20.Cf} {Interatomic potentials and forces}
     } 
} 
\maketitle

\section{Introduction}
\label{intro} 
Tremendous experimental progress has been made in the area of ultracold atomic systems over the past decades. 
The ability to dial the interactions among the particles via Feshbach resonances as well as the high-level of control over
the external confinement, have enabled to utilize these systems as simulation tools for various 
phenomena, from condensed-matter physics to quantum information \cite{bloch}. A recent ground-breaking achievement \cite{exp_ref,exp_ref2} is the ability to further confine 
just a few atoms in nearly isolated sites of optical lattices in which 
 atoms are cooled down to extremely low temperatures below the millikelvin. In the limit of low
tunneling, each lattice site may be regarded as an independent harmonic oscillator (HO) well. Theoretical 
models of many-body quantum systems that were considered idealizations in the past  can now be prepared
 experimentally and studied. For instance, the
limits of infinitely strong interactions  have recently been reached experimentally \cite{inns}
in one-dimensional (1D) geometry  using confinement-induced resonance \cite{conf}.
Under these extreme conditions, systems of bosons or distinguishable fermions,
display behavior similar to identical-fermion ensembles. By providing well-defined, strongly 
interacting many-body systems, the study of ultracold atoms gives physicists an 
unprecedented opportunity to assess theoretical techniques that cross the boundaries of disciplines. 

In this paper, we explore one of these strongly interacting  systems, namely an ensemble of two-component
 fermions confined in a HO potential. The temperature and density of matter reached in these 
ultracold dilute systems are such that the collisions among particles occur at small energy and the average interparticle distance
is much larger than the range of the interaction, which in the case of neutral atoms, is given by the van der Waals length. 
As a consequence of this separation of scales, the interaction among the particles can be modeled
 by a zero-range potential, such as the pseudopotential \cite{huang} which depends on a few low-energy parameters, {\it e.g.}, the S-wave two-body scattering length $a_2$ and the effective 
range $r_2$. For the two-particle system in a HO trap, analytical 
solutions with the pseudopotential exist and are  given by the so-called Busch formula \cite{busch,HT2,HT3}. 
Predictions of this exact model have been tested against experimental data and found to provide
 an accurate description of the two-atom system  \cite{exp_ref2,nikola_ref}. For systems with a larger
 number of particles, besides the three-body system at unitarity where analytical solutions exist \cite{werner}, one has to rely on numerical calculations.

In this work, we report on a new numerical study of few-fermion systems in a HO potential. The solution of the 
few-body problem is obtained by solving the Schr\"odinger equation in a truncated HO basis using the formalism of
 the No-Core Shell Model (NCSM) \cite{lee_nuc_phys1}. Since we work in a finite basis, an 
effective interaction has to be used instead of the pseudopotential which is only meaningful in the infinite Hilbert space.
This effective two-body interaction is generated with a unitary transformation of the exact two-body spectrum given by the 
Busch formula.  By construction the exact solutions of the two-particle system are reproduced in the truncated space. Unitary transformations have been widely used in different fields 
of physics. In Nuclear Physics effective nucleon-interactions have often been constructed using the Lee-Suzuki method \cite{lee_nuc_phys1,lee_nuc_phys2,lee_nuc_phys3}. 
Starting from a "bare" nucleon-nucleon potential defined in the infinite Hilbert space,
 the two-nucleon Schr\"odinger equation  is first solved
 within a as-large-as-possible (but finite) two-body model space assumed 
to resemble the infinite Hilbert space. The Lee-Suzuki transformation 
is then applied so that the interaction in  the truncated space reproduces
eigenenergies (usually the lowest) obtained with the bare interaction in the large two-body model space. 
In this study of HO-trapped fermions, the unitary transformation will be directly applied to the exact analytical
spectrum  given by the Busch formula.
Using this interaction, we will 
show that the energies of A-fermion systems (we will consider A=3,4,5) are in excellent agreement with results obtained 
by other methods.

Several other numerical studies of few-fermion systems in a HO trap have been  previously reported in the literature. Different approaches based on Monte Carlo technique 
such as, the Green's function Monte Carlo \cite{gfmc}, the fixed-node diffusion Monte Carlo \cite{blume} and the Shell-Model Monte Carlo \cite{nikola} methods have been 
applied to these systems. In \cite{trap0,us,us4,us5}, the NCSM formalism has been used to study few-fermion systems in a HO trap with an effective interaction 
constructed as a series of contact interactions and  derivatives of delta functions, according to the principles of Effective Field Theory \cite{bira,beda,plat}. Using 
the ground and few excited-states of the exact two-particle spectrum, a two-body effective interaction tailored to 
the truncation of the model space,  is constructed order by order in a systematic manner.
In \cite{alha},  the many-body problem is solved with the Configuration Interaction (CI)  to study systems in the unitary limit with an effective interaction 
 whose parameters are fixed such that the two-body eigenenergies given 
by this interaction correspond to the energies in the two-body exact spectrum. 
Our construction of the interaction has some similarity with that approach in the sense
that, in both cases, all the energies of the two-body spectrum obtained with the effective interaction, correspond
to the exact solutions given by the Busch formula. The authors in \cite{alha} assume a separable form for the two-body interaction whereas in
 our case, no such assumption is made. Moreover, we will consider not only the unitary limit but also  finite values of 
the scattering length and the effective range.

The paper is organized as follows. In  Sec. \ref{sec_theo} we recall some 
generalities about the two-particle system in a HO trap interacting with a short-range interaction and we show 
the construction of the effective two-body interaction.  In the following section, Sec. \ref{sec_res}, we show results
 for systems of three, four and five fermions and compare with results obtained by other methods. We conclude 
and discuss future applications in Sec. \ref{conclusion}.

\section{Formalism}
\label{sec_theo}
We consider a system of A identical fermions of mass $m$ interacting in a HO trap of frequency $\omega$.  For definiteness, we think of two-component fermions 
but the framework can be straightforwardly applied to other fermions as well as bosons.
The Hamiltonian $H_{lab}$ for this system in the laboratory frame reads:
\begin{equation}
H_{lab}=\sum_{i=1}^{A}\frac{{\vec{p}}_i^{~2}}{2 m}+\frac{m\omega^2\vec{r}_i^{~2}}{2}+\sum_{i<j}V_{ij} , \label{hami}
\end{equation}
where  $\vec{r_i}$ ($\vec{p_i}$) is the position (momentum) of particle $i$ with respect to the origin of the HO
potential (fixed in the laboratory) and $V_{ij}$ the two-body interaction  between particles $i$ and $j$. 
A nice property of the HO potential is the decomposition into a piece acting
on the center of mass (CM) of the A particles and a piece acting on their relative coordinates. Removing the CM part, we 
obtain  the Hamiltonian $H$ describing the intrinsic motion of the particles
\begin{eqnarray}
H=H_0+\sum_{i<j}V_{ij} , \label{hami_rel}
\end{eqnarray}
with
\begin{eqnarray}
H_0=\frac{1}{A}\left [ \sum_{i \ne j}\frac{(\vec{p_i}-\vec{p_j})^2}{2m} + \frac{ m \omega^2}{2}(\vec{r_i}-\vec{r_j})^2  \right ]. \label{hami_ho}
\end{eqnarray}
Using the formalism of the NCSM\cite{lee_nuc_phys1}, we  will diagonalize the Hamiltonian $H$ (\ref{hami_rel}) 
in a finite basis constructed with the eigenfunctions of the HO potential. The two-body (effective) interaction $V_{ij}$
 in (\ref{hami_rel}) will be tailored to the truncation of the model space and constructed from the exact solutions of 
the two-particle system.

\subsection{Two-fermion effective interaction}
We begin first by recalling some generalities about systems interacting at low energy, and then we describe the construction of the fermion-fermion interaction for systems in a HO trap.

It is well known that at sufficiently low energy, a physical system can be described without the complete knowledge of 
the interactions between its constituents. In free space, for a system of two particle with a relative momentum $k$
and  interacting via a short-range potential, the details
of the interaction at short-distance $r\le R$, with $r$ being the interparticle separation and $R$ the range, are 
irrelevant to describe the physics at $k \ll 1/R$. Moreover, due to the 
presence of the centrifugal barrier, only the lowest partial waves contribute and the S-wave dominates.
In this low-energy regime, the S-wave phase shift $\delta_0(k)$ is given by the Effective Range Expansion (ERE):
\begin{equation}
k \, \cot \delta_0(k) =-\frac{1}{a_2}+\frac{r_2 }{2}k^{2}+ \ldots, \label{ERE}
\label{ERE}
\end{equation}
where 
$a_2$, $r_2$, $\ldots$ are, respectively, the scattering length, effective range, and higher ERE parameters 
 not shown explicitly. For energies close to the scattering threshold, the S-wave scattering length $a_2$ primarily determines 
the scattering of two particles, as it can be seen in Eq. (\ref{ERE}) when $k \rightarrow 0$. 
Similarly, for a few(many)-body system at sufficiently low-energy, in which interparticle distances are large compared to the range of 
the interaction, the scattering properties will also depend mainly on $a_2$ \cite{hammer}. Physical properties of 
systems that depend only on the scattering length are {\it universal}. Universality in physics usually 
refers to situations in which systems that are very different at short distances have identical long-distance
behavior. Two universal systems 
may have completely different internal structures and interactions,
but the low-energy behavior will be the same as long as they have the same scattering length. Corrections coming from
details of the interaction, such as the effective range $r_2$ in Eq. (\ref{ERE}) are called {\it non universal}.

Ignoring short-distance details, it is then possible (and often practical) to substitute the original potential acting in the inner region,
by the zero-range pseudopotential \cite{huang},
\begin{eqnarray}
V^{(\delta)}=-\frac{2 \pi\hbar^2}{\mu}\frac{\tan(\delta_0(k)) }{k}\delta(\vec{r})\frac{\partial}{\partial r}r, \label{pseudo}
\end{eqnarray}
where $\mu$ is the reduced mass of the two-particle system.
By construction, solutions of the Schr\"odinger equation with the pseudopotential (\ref{pseudo}) have the same asymptotic behavior
as the original potential. The pseudopotential being of zero-range, the asymptotic behavior is reached for $r\rightarrow 0$
whereas for the original potential, it is reached beyond distance of order of the range  $R$. Hence, the wavefunctions given by 
both potentials differ only in the inner region. Equivalently, the solution of the Schr\"odinger equation with the pseudopotential (\ref{pseudo}) can be obtained by solving
the free Schr\"odinger equation supplemented by the following boundary condition at the origin,
\begin{eqnarray}
u(r)\rightarrow C \left (r+\frac{1}{k \cot \delta_0(k)} \right )  {~~~~~~~~\rm{for}~~~~} r~~\rightarrow 0, \label{boundary}
\end{eqnarray}
with $C$,  a normalization constant.

We now consider a HO potential of frequency $\omega$ in which two particles are trapped and interact via a short-range interaction. As in free space, we assume 
that the interaction between them  can be replaced by the pseudopotential (\ref{pseudo}), or 
equivalently, by the boundary condition (\ref{boundary}). The HO potential in the relative frame is given by  $V^{HO}(r)=\frac{1}{2}\mu \omega^2r^2$.  For $r>0$, 
 the S-wave wavefunction $\phi(r)$  of the two-particle system is solution of the Schr\"odinger equation 
\begin{equation}
\left ( \frac{-\hbar^2}{2\mu}\frac{d^2}{dr^2}+\frac{1}{2}\mu \omega^2r^2 \right )\phi(r)=E\phi(r) {~~~~~~~~~~\rm{for}~~~~~~} r>0. \label{hami_ho3}
\end{equation}
At the origin, the HO potential $V^{HO}(r)$  vanishes and $\phi(r)$  fulfills the boundary condition (\ref{boundary}). Solutions of Eq. (\ref{hami_ho3}) 
can be written with the Tricomi confluent hypergeometric function $U$ \cite{abram} as
\begin{equation}
\phi(r)=Are^{-\frac{r^2}{2b^2}}U \left (3/4-E/2\hbar \omega,3/2,r^2/b^2 \right ) \label{sol}
\end{equation}
where $A$ is a normalization constant and $b=\sqrt{\frac  {\hbar}{\mu \omega}}$ the HO length. 
From the behavior of the Tricomi function near the origin, one obtains
\begin{eqnarray}
\phi(r) \propto  \left ( 2 \frac{{\rm \Gamma}(-\nu)}{{\rm \Gamma}(-\nu-1/2)} r -b \right ) ~~~~{\rm for~~~}r\rightarrow 0 . \label{wf_zero}
\end{eqnarray}
where we use the notation $\nu=E/2\hbar \omega-3/4$.  
By matching the behavior of $\phi(r)$ at the origin (see Eq. (\ref{wf_zero})) with the boundary condition (\ref{boundary}), one obtains the Busch formula \cite{busch,HT2,HT3}: 
\begin{eqnarray}
\frac{{ \rm \Gamma}\left (3/4-E/2\hbar\omega \right )}{{\rm  \Gamma} \left (1/4-E/2\hbar \omega \right )}=-\frac{b}{2}k \cot \delta_0(k). \label{spectrum}
\end{eqnarray}
The above equation relates 
the energy of the two particles in the HO trap to the S-wave phase shift $\delta_0(k)$ in free space. Using the ERE (\ref{ERE}), one can rewrite Eq. (\ref{spectrum}) as: 
\begin{eqnarray}
\frac{{\rm \Gamma}\left (3/4-E/2\hbar \omega \right )}{{\rm \Gamma} \left (1/4-E/2\hbar \omega \right )}=\frac{b}{2a_2}-\frac{br_2}{4}k^2+\ldots \label{spectrum_ERE}
\end{eqnarray}
The Busch formula (\ref{spectrum_ERE}) was initially 
demonstrated in \cite{busch}  for interactions characterized with the ERE truncated to the scattering length $a_2$
and was latter generalized to include more terms from the ERE \cite{HT2,HT3}.

At this point, a remark should be made about the validity, for a trapped system, of 
replacing the original potential by  the pseudopotential (\ref{pseudo}) (or equivalently of using the boundary condition (\ref{boundary})). As we 
wrote previously, the pseudopotential 
 substitutes an interaction of short-range $R$ by a zero-range interaction that gives the same asymptotic behavior. If the HO
trap has a too large contribution in the inner region  of the original potential, the matching with asymptotic continuum wavefunction looses its meaning since particles cannot be considered free  anymore. As a consequence, the HO 
should be small in this region, that is, $b$ should be large enough {\it i.e.} $b/R \gg 1$ so that the matching procedure is not spoiled by the HO potential.  A study on the limitation of using 
the pseudopotential for a trapped system  can be found, for  instance,  in \cite{HT2}.

We now describe the construction of the effective two-body interaction.
Using the eigenenergy solutions of the Busch formula (\ref{spectrum}) and the corresponding wavefunctions (\ref{sol}),
 one can write the Hamiltonian $H^{(2)}$ of the two-particle system in the trap as
\begin{eqnarray}
H^{(2)}= U^{\dagger}E^{(2)}U \label{h_mat},
\end{eqnarray}
where $E^{(2)}$ is the diagonal matrix formed with the eigenenergies and  $U$ is the matrix formed with the
associated eigenvectors\footnote{For each eigenenergy, the eigenvector can easily be expressed
 in any basis from the corresponding radial wavefunction (\ref{sol}).}. 
The Hamiltonian $H^{(2)}$ is defined in the infinite Hilbert space.
An obvious choice of a basis for this space is the set of relative 
S-wave HO states $|n\rangle$ characterized by the HO radial quantum number $n$ and the energy $E_n=(2n+3/2) \hbar \omega$. 
Introducing the parameter $n_{\rm{\rm{max}}}$, we perform the following partition of the two-particle space: 
the HO states with radial quantum number smaller or equal to $n_{\rm{max}}$ define the model space ${\cal P}$, within which the effective interaction will be constructed. 
 The remaining states, {\it i.e.}, the states with radial quantum larger than $n_{\rm{max}}$, define the complementary space ${\cal Q}$. 
The largest number of quanta $N_2^{\rm{max}}$, carried by states of ${\cal P}$ is  $N_2^{\rm{max}}=2n_{\rm{max}}$. 
Introducing the projectors $P$ and $Q$ acting respectively  in ${\cal P}$  and ${\cal Q}$,
the matrix $U$ in Eq. (\ref{h_mat}) can be divided into four blocks: 
\begin{eqnarray}
 \left( \begin{array}{cc}
U_{PP} & U_{PQ} \\
U_{QP}  & U_{QQ} \end{array} \right) 
\end{eqnarray}
Each block is obtained by application of the projectors on $U$: for instance, by definition $U_{PQ}=PUQ$.
Similarly, $E^{(2)}$ can be split as
\begin{eqnarray}
 \left( \begin{array}{cc}
E^{(2)}_{PP} & 0 \\
0  & E^{(2)}_{QQ} \end{array} \right) 
\end{eqnarray}
where $E^{(2)}_{PP}$ ($E^{(2)}_{QQ}$) is the diagonal matrix formed  by the eigenenergies Eq. (\ref{spectrum}) restricted to ${\cal P}$ (${\cal Q}$).
The effective Hamiltonian $H^{\rm{eff}}_{P}$ is constructed such that the energies in the truncated model space, {\it i.e.} ${\cal P}$, correspond to the lowest energies of the original Hamiltonian  $H^{(2)}$. We construct $H^{\rm{eff}}_{P}$ with the same  unitary transformation as in \cite{alex}:
\begin{eqnarray}
H^{\rm{eff}}_{P}=\frac{U_{PP}^{\dagger}}{\sqrt{(U_{PP}^{\dagger}U_{PP})}}E^{(2)}_{PP}\frac{U_{PP}}{\sqrt{(U_{PP}^{\dagger}U_{PP})}}. \label{hami_eff}
\end{eqnarray}
The  two-body effective interaction  $V^{\rm{eff}}_P$ acting in $P$ 
is obtained by subtracting the HO potential to the effective Hamiltonian $H_P^{\rm{eff}}$:
\begin{eqnarray}
V^{\rm{eff}}_P=H^{\rm{eff}}_{P}-PH_0P. \label{V_eff}
\end{eqnarray}
Let us consider the limit of $H^{{\rm{eff}}}_P$ as the model space ${\cal P}$ goes to the full Hilbert space {\it i.e.} as $n_{\rm{max}} \rightarrow \infty$. In this limit,  the matrix product $U^{\dagger}_{PP}U_{PP}$ will converge
to the Identity operator for systems with vanishing $r_2$ and higher order terms in the ERE. The effective 
Hamiltonian $H^{(\rm{eff})}_P$ will then converge to the original Hamiltonian $H^{(2)}$ (see Eq. (\ref{hami_eff})). 
In the case of finite $r_2$,  the eigensolutions of the Busch formula (\ref{spectrum_ERE}) are energy-dependent and the eigenstates are not orthogonal to each other. As a consequence, as ${\cal P}$ becomes the full Hilbert space,
 $U^{\dagger}_{PP}U_{PP}$ converges  to  a different operator than the Identity. Nevertheless, by construction, the eigenvectors of  the effective Hamiltonian $H^{\rm{eff}}_{P}$ (Eq. \ref{hami_eff}) will still be  orthogonal to each other.

We thus have designed a two-body effective interaction $V^{\rm{eff}}_P$ that reproduces the lowest
 energies of the two-particle system in the trap given by the Busch formula. From now on, we write $V^{\rm{eff}}_{n_{\rm{max}}}$ the interaction
in the two-body space characterized by the cutoff $N_2^{\rm{max}}=2n_{\rm{max}}$. 

We briefly describe in the following, the construction of the NCSM basis for the resolution of the many-body Schr\"odinger equation (\ref{hami_rel}).

\subsection{Many-body basis}
\label{many_sec}
The resolution of the Schr\"odinger equation for the few-fermion system in a HO trap comes down now to the diagonalization 
of the effective Hamiltonian $H^{(\rm{eff})}$:  
\begin{eqnarray}
H^{(\rm{eff})}=H_0+\sum_{i<j}V^{\rm{eff}}_{n_{\rm{max}},ij}, \label{hami_rel_eff}
\end{eqnarray}
where $V^{\rm{eff}}_{n_{\rm{max}}, ij}$ is the effective two-body interaction 
between particles $i$ and $j$. 
The above Hamiltonian is diagonalized using the formalism of the NCSM where the basis is constructed from  HO wavefunctions. 
Here we work with 
 Jacobi coordinates \cite{petr} defined  in terms of differences between the CM positions of 
sub-clusters within the $A$-body system, e.g.,
\begin{eqnarray}
\vec{\xi}_1&=&\sqrt{\frac{1}{2}}\left(\vec{r}_1-\vec{r}_2\right),
\nonumber\\
\vec{\xi}_2&=&\sqrt{\frac{2}{3}}
\left[\frac{1}{2}\left(\vec{r}_1+\vec{r}_2\right)-\vec{r}_3\right],
\nonumber\\
\vdots && 
\nonumber\\
\vec{\xi}_{A-1}&=& \sqrt{\frac{A-1}{A}}
\left[\frac{1}{A-1}\left(\vec{r}_1+\cdots+\vec{r}_{A-1}\right)
-\vec{r}_{A}\right] .
\end{eqnarray}
The basis states are eigenstates of the HO potential $H_0$ (\ref{hami_ho}) expressed in Jacobi coordinates.  
For the purpose of illustration, let us consider a system of A=3 fermions of spin $s=1/2$. In that case 
two Jacobi coordinates $\vec{\xi}_1, \vec{\xi}_2$ are necessary to describe the intrinsic motion and the antisymmetrized basis
states can be written as:
\begin{eqnarray}
{\cal A}\left\{\left[\phi_{n l}(\vec \xi_1) \otimes \phi_{{\cal NL}}(\vec \xi_2) 
\right]_{L} |(s s ){\cal S}s;S\rangle\right\} \label{state}
\end{eqnarray}
where $n,l$ ($N, {\cal L}$) are respectively the radial quantum number and angular momentum associated with the Jacobi coordinate $\vec{\xi}_1$ ($\vec{\xi}_2$). The angular momenta $l$ and ${\cal L}$ are coupled to the total angular momentum $L$ whereas the three spins $s$ are coupled to total spin S. 
The antisymmetry is enforced by the operator ${\cal A}$; more details on the antisymmetrization can be found in \cite{petr}.
The state (\ref{state}) is an eigenstate of the HO Hamiltonian with the energy  $(N_3+3)\hbar \omega$ where $N_3$ is the quantum
number $N_3$ defined by $N_3 = 2n + l + 2\cal{N} + \cal{L}$. For a system of A particles, the model space is truncated by introducing a cutoff $N_A^{\rm{max}}$ defined as the 
largest number of quanta in the eigenstates of HO used to construct the A-body basis. 
Using again the three-body system as a concrete example, truncating the three-body basis at $N^{\rm{max}}_3$ means keeping only states
with HO energies such that $N_3 = 2n + l + 2N + L \leq N^{\rm{max}}_{3}$. 

Instead of working with Jacobi coordinates, we could have chosen to construct the NCSM basis as a set of Slater determinants
 from single-particle HO wavefunctions defined in the laboratory frame. The antisymmetry is easily built into this approach but the dimension of the basis space grows quickly with $N^{\rm{max}}_{A}$.
On the contrary, working with Jacobi coordinates is more 
effective than a Slater-determinant basis in the sense that the dimension 
of the model space grows more smoothly with $N^{\rm{max}}_A$. On the other hand, with this choice of coordinates, the antisymmetrization becomes increasingly difficult as the number of particles grows \cite{petr}. But, since in the current paper, we 
investigate systems with at most five fermions, we have chosen the Jacobi coordinates 
approach in order to be able to consider large $N^{\rm{max}}_A$.

We show in the following results for the energy of few-fermion systems in a HO trap for various values of 
the scattering length $a_2$ and effective range $r_2$. 

\section{Results}
\label{sec_res}
We now present numerical results for systems of few two-component fermions interacting in a HO trap. The effective 
two-fermion interaction is constructed as described in the previous section.

\subsection{Three fermions at unitarity}
We begin here with an ensemble of three fermions at unitarity {\it i.e.}  $\frac{a_2}{b}=\infty$ and $\frac{r_2}{b}=0$. 
This systems gives us a perfect ground for testing our numerical approach since semi-analytical solutions exist in that case \cite{werner}. In this regime, the 
only length scale characteristic of the system is given by the HO radius $b$. 
The energies of the two-fermion system at unitarity are solutions of the Busch formula  (\ref{spectrum_ERE})  and
are  given by
\begin{eqnarray}
E=(1/2+2n)\hbar\omega \label{sp_uni}
\end{eqnarray}
where $n$ is integer such that $n \ge 0$. The two-body interaction $V^{\rm{eff}}_{n_{\rm{max}}}$ 
is constructed  in the truncated two-body model space characterized by the cutoff $N_2^{\rm{max}}=2n_{\rm{max}}$ 
and  by construction, $V^{\rm{eff}}_{n_{\rm{max}}}$ reproduces the  $n_{\rm{max}}+1$ lowest energies in the spectrum (\ref{sp_uni}).
We study the convergence of the three-body energy as the cutoff $N_2^{\rm{max}}$ is increased.
The three-body NCSM basis is characterized by the parameter $N_3^{\rm{max}}$ defined as the largest number of HO quanta carried by 
the states forming this basis. For each values of $N_2^{\rm{max}}$, the size of the three-body model space is increased until a convergence is reached \cite{us4,us5,alha}. 

\begin{figure}
\rotatebox{-90}{\resizebox{0.32\textwidth}{!}{%
  \includegraphics*{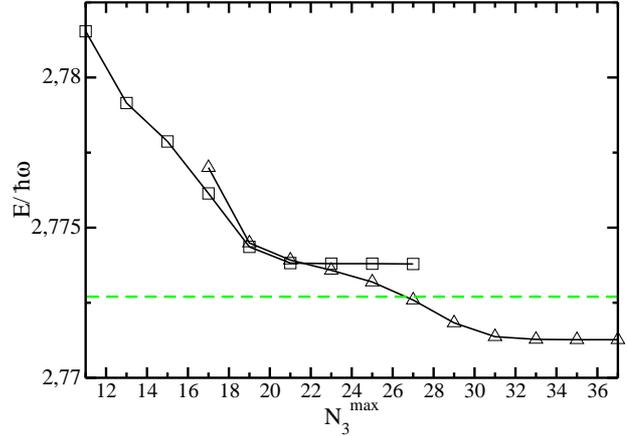}}}
\caption{Energy in units of the HO frequency, $E/\hbar\omega$, of the ground-state $L^{\pi} = 1^{-}$ of the A = 3
system at unitarity, as function of the three-body model-space size, $N^{\rm{max}}_3$: $N^{\rm{max}}_2=10$ (square) and $N^{\rm{max}}_2=16$ (triangle). The dashed line marks the exact value \cite{werner}.}
\label{fig1}
\end{figure}

\begin{figure}[tb]
\rotatebox{-90}{\resizebox{0.32\textwidth}{!}{%
  \includegraphics*{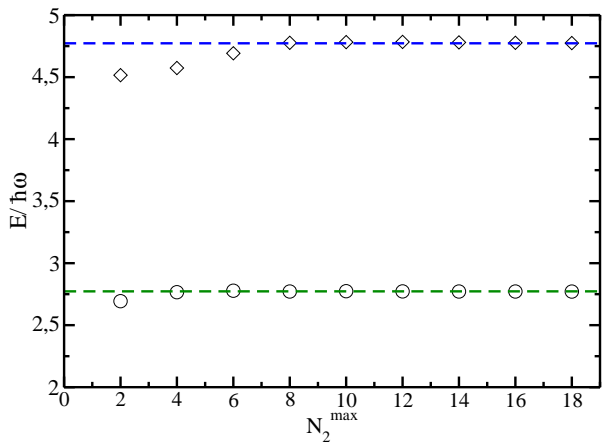}}}
\caption{Energy in units of the HO frequency, $\frac{ E}{\hbar \omega}$, of the ground state (circle) and first excited state (diamond) coupled to  $L^{\pi} = 1^{-}$ of the A = 3
system at unitarity, as function of the two-body cutoff, $N_2^{\rm{max}}$. The dashed lines mark the exact values for the ground state and the first excited state \cite{werner}.}
\label{states_3_unitarity}
\end{figure}

The ground-state solution of the three-fermion system is coupled to  $L^{\pi}=1^{-}$ and we show in  Fig. \ref{fig1}
its energy for $N_2^{\rm{max}}=10, 16$ as a function of $N^{\rm{max}}_3$. For a fixed cutoff $N^{\rm{max}}_2$, the Hamiltonian is fixed 
and one expects a variational behavior of the energy as the three-body model space is increased. This can 
be seen in Fig. \ref{fig1} where  for each  value of  $N^{\rm{max}}_2$, the energy decreases sharply first as $N^{\rm{max}}_3$ 
increases, until a convergence to a final value is reached. The value of $N^{\rm{max}}_3$ for which the
convergence is obtained depends on the particular value of $N^{\rm{max}}_2$. For  $N^{\rm{max}}_2=10$,  the
energy of the three-body ground state does not change by more
than $10^{-4}$ (in units of $\hbar\omega$) when $N_3^{\rm{max}} \ge 21$ while for  $N^{\rm{max}}_2=16$ convergence at this level is achieved for  $N_3^{\rm{max}}\ge 33$ (see Fig. \ref{fig1}).
Figure \ref{states_3_unitarity} shows the convergence of the ground-state energy as a function of $N^{\rm{max}}_2$. 
The agreement with the exact solution is very good: for the largest cutoff we have considered here, $N_2^{\rm{max}}=20$, we obtain $2.7712$ for the ground-state energy (in units of $\hbar\omega$)  as compared to the exact solution $2.7727$ \cite{werner}. Moreover, the 
convergence with respect to $N^{\rm{max}}_2$ is rather fast. Indeed,
starting at $N_2^{\rm{max}}=4$, all the circles representing the energy of 
the ground state lie on top of the exact result represented by the dashed-line (see Fig. \ref{states_3_unitarity}).
At $N_2^{\rm{max}}=4$, the energy obtained is  $2.7657$ which corresponds 
to a difference with the exact result smaller than 0.3 \%.

The calculation of excited-state energies can be carried out in a similar fashion. Again, for each value of $N_2^{\rm{max}}$, we increase $N_3^{\rm{max}}$ until convergence at $10^{-4}$ (in units of $\hbar\omega$) is reached. For the first-excited state coupled to $L^{\pi} = 1^{-}$, convergence at a fixed  $N_2^{\rm{max}}$ is reached for larger values of $N_3^{\rm{max}}$ as compared to the ground state: for $N_{2}^{\rm{max}}=10$ and  $N_{2}^{\rm{max}}=16$, one has to go to respectively to $N_3 \ge 23 $ and $N_3 \ge 35 $ 
to reach the same precision. Results for the first excited-state energy as a function of $N_2^{\rm{max}}$ are shown in  Fig. \ref{states_3_unitarity}. As for the ground state,  the 
agreement with the exact value is excellent: for $N_{2}^{\rm{max}}=20$, the energy is equal to $4.7732$ as compared  to the exact value $4.7727$ \cite{werner}. 
Relatively to the ground state, the first-excited energy converges more slowly with $N^{\rm{max}}_2$ (see Fig. \ref{states_3_unitarity}). But still, the energy 
approaches rapidly the exact solution: already at $N_2^{\rm{max}}=6$, the energy is $4.7769$ which corresponds to a difference with the exact value $4.7727$   smaller than 0.1 \%.

\begin{figure}[tb]
\rotatebox{-90}{\resizebox{0.32\textwidth}{!}{%
  \includegraphics*{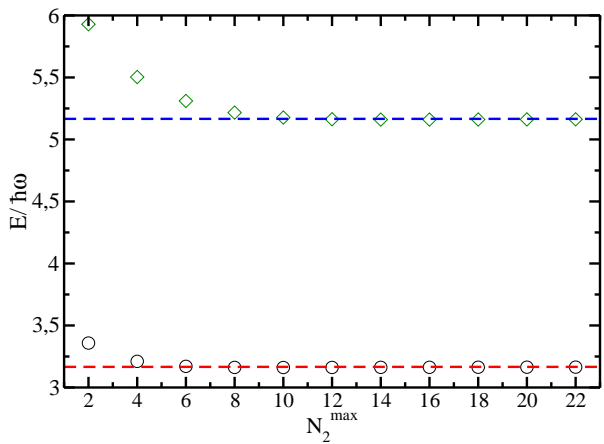}}}
\caption{ Energy in units of the HO frequency, $E/\hbar \omega$, of the ground state (circle)and first excited state (diamond) coupled to $L^{\pi} = 0^+$ of the A = 3
system at unitarity, as function of the two-body cutoff, $N_2^{\rm{max}}$. The dashed lines mark the exact value \cite{werner}.}
\label{states_3_unitarity_zero}
\end{figure}

We have also studied states with other couplings. The general features of convergence with $N_2^{\rm{max}}$  and $N_{3}^{\rm{max}}$ are similar as  previously. We show in Fig. \ref{states_3_unitarity_zero}, the results for the ground state and first-excited state coupled to $L^{\pi} = 0^+$ as a function of $N_{2}^{\rm{max}}$.  The convergence with respect to $N_2^{\rm{max}}$  is fast and in excellent agreement with the exact values: for the ground (first-excited) state, at $N_{2}^{\rm{max}}=22$, the energy is equal to 3.1640 (5.1622) whereas the exact energy is 3.1662 (5.1662) \cite{werner}. 
\subsection{Finite scattering length and effective range}
The construction of the two-body effective interaction is general and allows us to consider also physics away from unitarity. As one moves 
away from a Feshbach resonance, the effects of the interaction range should become more pronounced and universality reduced to some degree. Here, we will 
assume the following values for the scattering length and the effective range: $b/a_2=1, r_2/b=0.1$. This choice is motivated by the 
fact that this system has been studied in \cite{us4} using interactions constructed from the EFT principles, and allows us to test our method away from unitarity.

\begin{figure}[h!]
\rotatebox{-90}{\resizebox{0.32\textwidth}{!}{%
  \includegraphics*{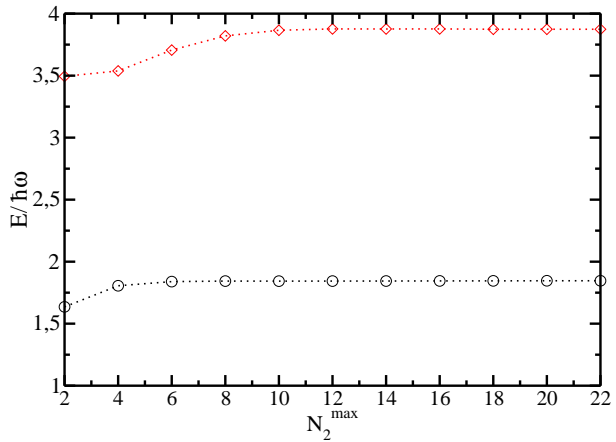}}}
\caption{ Energy in units of the HO frequency, $E/\hbar \omega$, of the ground state (circle) and first excited state (diamond) coupled to $L^{\pi} = 1^-$ of the A = 3
system for $b/a_2=1, r_2/b=0.1$ as function of the two-body cutoff, $N_2^{\rm{max}}$. The dotted lines are drawn to guide the eye.}
\label{states_3_finite}
\end{figure}

Figure \ref{states_3_finite} shows the energy of the ground state and the first excited 
state coupled to $L^{\pi}=1^-$ as a function of $N_2^{\rm{max}}$. As previously in the unitary regime, for each value of $N_2^{\rm{max}}$, the three-body model 
space is increased until convergence at $10^{-4}$ is reached. Again,  the convergence with $N_2^{\rm{max}}$ is fast: for the largest cutoff we have considered, $N^{\rm{max}}_2=20$, the energy for the ground state 
and the first excited state  are respectively  $1.8458$ and $3.8739$. Our results agree nicely with the energies obtained in the EFT approach: at N$^2$LO,  the ground-state and first-excited state energy were respectively found at $1.8443$ 
and $3.8680$ \cite{us4}.

\subsection{Four- and five-fermion systems  at  unitarity}

We now consider heavier systems with  A=4 and A=5 fermions. We show only results at unitarity
 although finite values for the scattering length and effective range could be considered as well.

Again for each $N_2^{\rm{max}}$,  the Schr\"odinger equation is solved in a truncated many-body model space whose size $N^{\rm{max}}_A$
is increased  until convergence is reached.  Because of the larger number of particles, we limit ourselves to smaller space.

\begin{figure}[tb]
\rotatebox{-90}{\resizebox{0.32\textwidth}{!}{%
  \includegraphics*{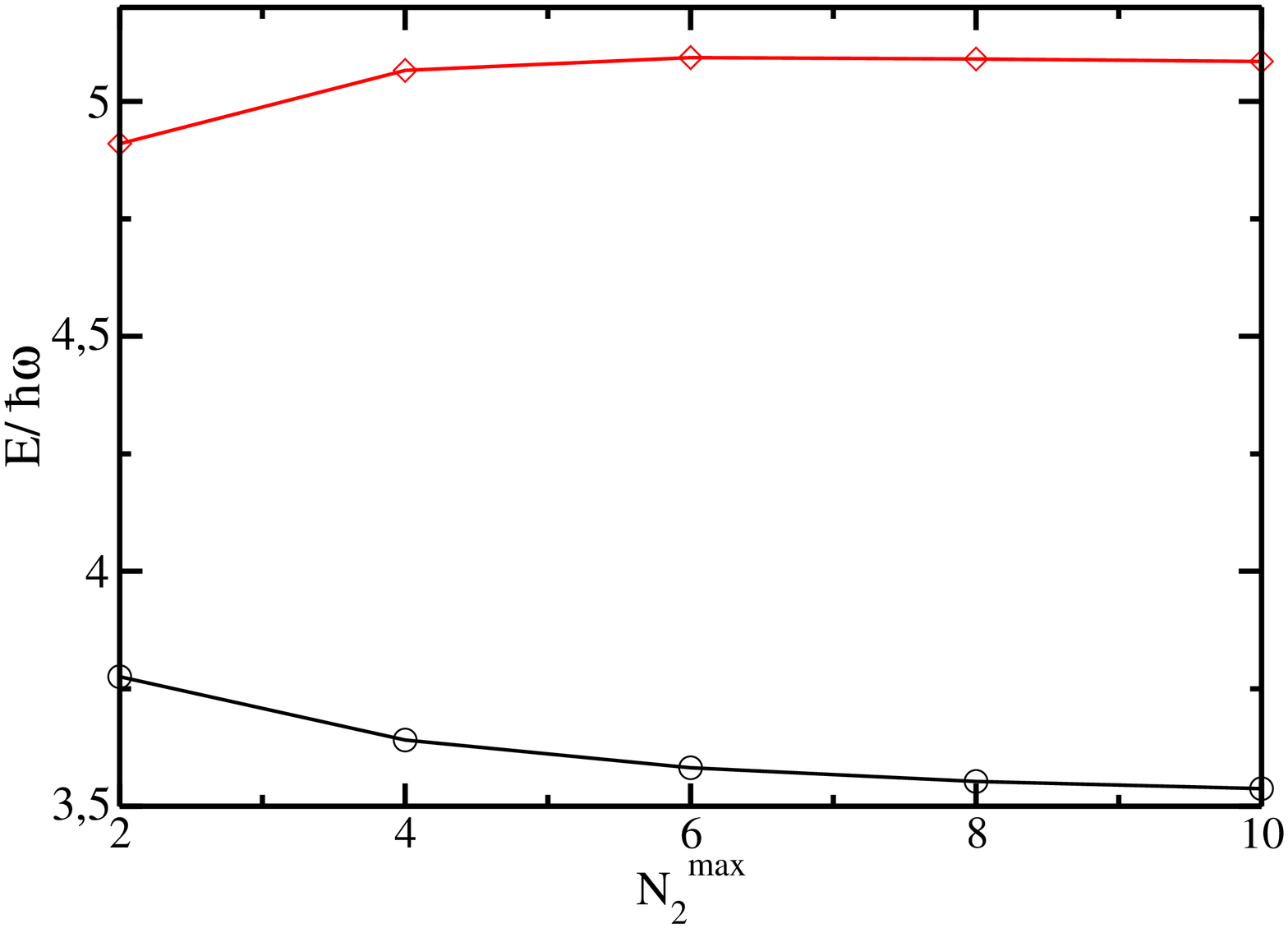}}}
\caption{Energy in units of the HO frequency, $E/\hbar \omega$, of the ground state (circle)and first excited state (diamond) coupled to $L^{\pi} = 0^+$ of the A = 4
system at unitarity as function of the two-body cutoff, $N_2^{\rm{max}}$.}
\label{states_four}
\end{figure}

Figure \ref{states_four} shows the ground-state and first-excited state energy of the four-body system  coupled 
to $L^{\pi}=0^{+}$ as a function of $N_2^{\rm{max}}$. The value $N_4^{\rm{max}}$ has been increased for each $N_2^{\rm{max}}$ until the energies do not change by  more than $10^{-3}$.
%
For the largest cutoff considered here, $N_2^{\rm{max}}=10$, we obtain a ground-state energy equal to  3.537. This result is in good agreement 
with calculations  using EFT interaction at N$^2$LO where the energy was found to be 3.52 \cite{us4}, and also with results from various different approaches, where 
the ground-state energy was found to be 3.509 \cite{stecher}, 3.6 $\pm$ 0.1 \cite{gfmc}, 3.551 $\pm$0.009 \cite{blume}, 
and 3.545 $\pm$ 0.003 \cite{alha}. 
For the first-excited state we obtain $5.085$ which is also in good agreement 
with  EFT results at N$^2$LO where the energy was found to be 5.07 \cite{us4}.
\begin{figure}[tb]
\rotatebox{-90}{\resizebox{0.32\textwidth}{!}{%
  \includegraphics*{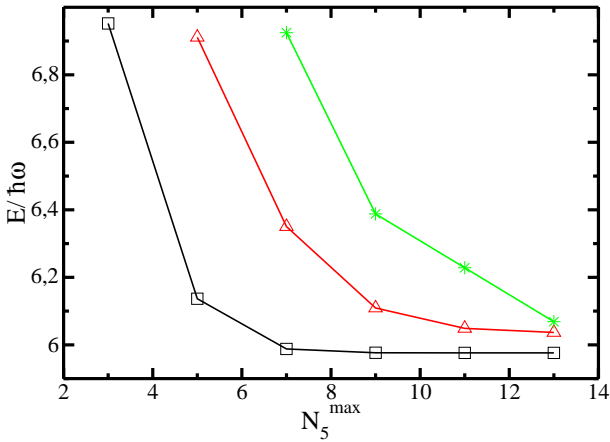}}}
\caption{Energy in units of the HO frequency, $E/\hbar \omega$, of the ground state 
 of the A = 5 system at unitarity, as function of the five-body model-space size, $N^{\rm{max}}_5$: $N^{\rm{max}}_2=2$ (square), $N^{\rm{max}}_2=4$ (triangle) and $N^{\rm{max}}_2=6$ (star).}
\label{five_fig}
\end{figure}

Finally, we present results for the ground state energy of the five-fermion system at unitarity. Figure \ref{five_fig} shows the ground-state energy
for several cutoff $N_2^{\rm{max}}$, as a function of $N_5^{\rm{max}}$. As in previous cases, the energy  at fixed $N^{\rm{max}}_2$ decreases when the many-body model space
grows, and convergence is faster for smaller $N^{\rm{max}}_2$. For the largest five-fermion model space we consider here, {\it i.e.} for $N^{\rm{max}}_5=13$,
results have converged nicely at $N_2^{\rm{max}}=2$ and  $N_2^{\rm{max}}=4$. For this latter cutoff, the difference between the energy at $N_5^{\rm{max}}=11$ and  $N_5^{\rm{max}}=13$
is $\sim 1.2~10^{-2}$, which represents less than 0.2 \% of the energy 6.036 obtained at $N^{\rm{max}}_5=13$. For $N_2^{\rm{max}}=6$, the 
difference in the energy for the same values of $N_5^{\rm{max}}$ 
is, as expected, larger. It is equal in that case to $\sim 1.6~10^{-1}$ and this difference corresponds to  2.6 \% of the energy 6.068 obtained at $N^{\rm{max}}_5=13$. 
For all $N_2^{\rm{max}}$ we have considered, results at $N_5^{\rm{max}}=13$ are all within a small range of energy that goes from $5.976$ to $6.068$ (5.976 is the ground-state energy for $N_2^{\rm{max}}=2$). 
For the largest cutoffs $N^{\rm{max}}_2=4$ and $N^{\rm{max}}_2=6$, the results for the ground-state energy are respectively 6.036 and 6.068. These values are in good agreement
with the value $5.9565(15)$ obtained with the stochastic variational method in \cite{blume_CR} and with the energy $6.1(1)$ obtained with the GFMC approach \cite{gfmc}.  

As we previously mentionned in Sec. \ref{many_sec}, when working with Jacobi coordinates, the size of 
the model space grows smoothly with $N^{max}_{A}$.  The largest matrix we had to consider was for the
 A=5 system at $N^{max}_A=13$ where the dimension is equal to 4361. This a rather small diagonalization problem than can easily be handled numerically. On the other hand, in Jacobi coordinates, the antisymmetrization becomes increasingly difficult as the number
of particles grows \cite{petr}. For larger systems, it would be more efficient to work in a Slater determinants basis
constructed from single-particle HO wavefunctions in the laboratory frame. Nowadays, the largest ensembles 
that can be treated with NCSM codes in Slater Determinants basis are nuclear systems with $A \sim 10$
 where the matrix dimension can reach size of the order of $10^9$ \cite{sm_codes}.
\section{Conclusions and Outlook}
\label{conclusion}
We have studied systems of few fermions in a HO trap interacting via a
two-body effective interaction generated from a unitary transformation of the exact two-body spectrum. Constructions of interactions 
tailored to a truncated space using unitary transformations are common in many fields of physics. 
The originality of our approach comes from the fact that the unitary transformation is directly performed on the exact two-body spectrum which is independent
of the form of a initial "true" two-body potential. 
Indeed, the energies of the two-body system in the HO potential are given by the Busch formula as function of few model-independent, low-energy 
parameters, namely the S-wave scattering length $a_2$ and effective range $r_2$.  
This model independence is enabled by the separation of scales inherent to these systems. Indeed, the interparticle distance being
 much larger than the range of the interaction and the matter density much lower than the typical matter densities,
the physics is only sensitive to the long-distance (low-energy) components of the interaction which can be taken into account, in a model-independent fashion, by the ERE.
By construction, the two-body effective interaction reproduces, in the truncated space, the energies solution of the Busch formula.

Using this effective two-body interaction, we have considered several few-fermion systems and studied the convergence of their energies with $N^{\rm{max}}_2$
the interaction cutoff. The few-body Schr\"odinger equation has been solved using the formalism of the NCSM.
We first studied systems of $A$=3 particles at unitarity  and at finite values of  $a_2$, $r_2$.
We have showed an excellent agreement with the exact solutions at unitarity for the ground state and the first-excited state. For 
finite values of $a_2$ and $r_2$, we have seen a similar agreement with the results obtained by the  EFT approach of \cite{us4}. 
We then considered systems of $A$=4,5 fermions at unitarity and similarly, we have obtained an excellent
agreement for the energies with results obtained from several other numerical approaches \cite{gfmc,blume,us4,alha,stecher,blume_CR}.

The construction of the effective interaction from the Busch formula is quite general and can be used to study other systems such as few-fermion mixtures
consisting of different species, and bosonic systems. An interesting application could be 
the study of the evolution of an ensemble of few bosons in a 1D trap, from the weakly-interacting regime of Bose-Einstein Condensate 
to the strongly-repulsive regime of the Tonks-Girardeau gas \cite{tonk}. In this latter case, the 
infinite repulsion between bosons imposes an effective Pauli-principle, and the 
many-body wavefunction can be mapped to the wavefunction of non-interacting fermions. The crossover from the Tonks-Girardeau  gas to the strongly interacting phase {\it i.e.} the  super Tonks-Girardeau gas,
 has been very recently realized experimentally \cite{inns} and could also be studied with our approach.
Several theoretical techniques \cite{zol,deur,brou} have been considered to study these 1D systems. In the most recent one \cite{brou},
the authors have developed an elegant method where the many-body wavefunction is constructed from a correlated
pair wave function (CPWF). The CPWF is inspired from the idea that the analytical solution for a single pair in the HO trap can be extended to the many-body system
by taking products of exact two-body functions. By construction, this method reproduces the exact known solutions for many-body systems in the limits of
zero and infinite  interactions.  For the two-particle system, the exact value solution of the Busch formula for any arbitrary interaction is reproduced.
 The CPWF method has been applied to 1D bosonic systems and two-component few-fermions mixtures and has shown an overall good
agreement with other approaches, especially in regime close to the Tonks-Girardeau gas or weakly-interacting systems \cite{brou}.

Our unitary-transformation-based approach has some similarities with the CPWF. Indeed, by construction, the two-body spectrum given by the Busch formula is also exactly reproduced
in our case. It would then be interesting to compare results obtained by both methods for quantities such as
 the momentum distributions, the matter density  and the occupation  number in  the HO trap shells
 as the interaction evolves from the weak to the infinitely-strong regime. Work in that 
direction is currently underway \cite{jonathan}.

\begin{acknowledgement}
Discussions with C. Forss\'en and  N.T. Zinner are gratefully acknowledged. We thank Petr Navr\'atil for providing
a Jacobi coordinates code to solve the four- and five-body problem.
This research was supported in part 
by the European Research Council (ERC StG 240603) under the FP7,
the US NSF under grant PHY-0854912,
and the US DOE under grant DE-FG02-04ER41338.
\end{acknowledgement}

\end{document}